\documentclass[
 reprint,
 superscriptaddress,
 amsmath,amssymb,
 aps, prapplied,
 longbibliography
]{revtex4-2}
\usepackage{multirow} 
\usepackage{amsmath}
\usepackage[table,xcdraw]{xcolor}
\usepackage{graphicx}
\usepackage{dcolumn}
\usepackage{bm}
\usepackage{float}

\usepackage{algpseudocode}
\usepackage{algorithm}
\usepackage{physics}
\usepackage{color}
\usepackage{placeins}
\usepackage[colorlinks=true, linkcolor=blue, citecolor=blue, urlcolor=blue]{hyperref}

\usepackage[capitalise]{cleveref}
\usepackage[all]{hypcap}
\usepackage{booktabs} 
\usepackage{amsmath} 						
\usepackage{amssymb} 						
\usepackage{mathtools}			
\usepackage{mathrsfs}                       
\usepackage{amsthm}
\usepackage{amssymb}
\usepackage{makecell}
\usepackage[utf8]{inputenc}
\usepackage{soul}
\usepackage{xcolor}

\soulregister\cite7
\soulregister\ref7
\soulregister\pageref7

\usepackage{booktabs}  
\usepackage{color}
\usepackage{xcolor}

\begin{document}
\global\setlength{\abovedisplayskip}{6pt}
\global\setlength{\belowdisplayskip}{6pt}
\global\setlength{\abovedisplayshortskip}{3pt}
\global\setlength{\belowdisplayshortskip}{3pt}
\setlength{\abovecaptionskip}{2pt}   
\setlength{\belowcaptionskip}{3pt}   
\preprint{APS/123-QED}

\title[Author guidelines for IOP Publishing journals in  \LaTeXe]{Efficient time-series prediction on NISQ devices via time-delayed quantum extreme learning machine}


\author{Mio Kawanabe}
\affiliation{Department of Electrical Engineering and Computer Science,
Tokyo University of Agriculture and Technology, Koganei, Tokyo 184-8588, Japan}
\author{Saud Čindrak}
\affiliation{Institute of Physics, Technische Universit\"at Ilmenau,
Ilmenau, Germany}
\author{Kathy Lüdge}
\affiliation{Institute of Physics, Technische Universit\"at Ilmenau,
Ilmenau, Germany}
\author{Jun-ichi Shirakashi}
\email[Contact author: ]{shrakash@cc.tuat.ac.jp}
\affiliation{Department of Electrical Engineering and Computer Science,
Tokyo University of Agriculture and Technology, Koganei, Tokyo 184-8588, Japan}
\author{Tetsuo Shibuya}
\affiliation{Division of Medical Data Informatics, Human Genome Center,
The Institute of Medical Science, The University of Tokyo,
Minato, Tokyo 108-8639, Japan}
\author{Hiroshi Imai}
\affiliation{The Graduate School of Information Science and Technology,
The University of Tokyo, Bunkyo, Tokyo 113-8656, Japan}

\date{\today}

\begin{abstract}
We proposed a time-delayed quantum extreme learning machine (TD-QELM) for efficient time-series prediction on noisy intermediate-scale quantum (NISQ) devices. By encoding multiple past inputs simultaneously, TD-QELM achieves shallow circuit depth independent of sequence length, thereby, mitigating noise accumulation and reducing computational complexity. Experiments using the NARMA benchmark on both noiseless simulations and IBM's 127-qubit processor demonstrate that TD-QELM consistently outperforms conventional quantum reservoir computing in prediction accuracy and noise robustness. These results highlight TD-QELM as a practical and scalable framework for time-series learning on current NISQ hardware.
\end{abstract}

\maketitle

\textit{Introduction.} In recent years, quantum computing has emerged as a transformative technology with the potential to revolutionize a wide range of scientific and engineering disciplines \cite{NIE06a}.  However, current quantum computers remain limited by noise originating from imperfections in quantum hardware, placing them in the era of noisy intermediate-scale quantum (NISQ) devices {\cite{ABU25, PRE18}}. Quantum algorithms that promise computational advantage typically require millions of qubits and deep circuits, rendering them impractical for near-term devices.  

Among the various approaches compatible with NISQ hardware, \emph{quantum machine learning} (QML) has emerged as a promising and robust framework \cite{ARU19, SCH15a, BIA17, CER22}. QML explores how machine learning algorithms can be efficiently implemented using quantum systems. Within this field, \emph{quantum reservoir computing} (QRC) employs quantum systems as dynamical reservoirs in which only the readout layer is trained \cite{FUJ17}. 
This approach has attracted significant attention, particularly in the domains of time-series forecasting and nonlinear dynamical modeling \cite{CHE20b, SUZ22, NAK19, GAR23, MUJ21a, ABB24a, FUJ21, NAK20, PFE23, GOE25,VET25}. Recent research show cases that optimal performance is achieved at dynamical phase transitions \cite{MAR21} and at the edge of quantum chaos \cite{KOB26}. Furthermore, \cite{CIN25, CIN25a} introduced a measure based on Krylov complexity, showing near-perfect correlations with data expressivity in QRC, thus linking computational performance with insights into complex quantum systems.

Although noise is often detrimental in quantum algorithms, specific types of noise have been shown to enhance computational performance in QRC \cite{CHE20b, SUZ22, KUB23, DOM23, SAN24, MON25, FRA24, FRY23}. For instance, dissipative QRC has been shown to outperform noiseless implementations \cite{CHE20b, SAN24, DOM23, FRY23}. In a similar vein, Suzuki et al. demonstrated that the intrinsic noise inherent to superconducting quantum processors can enhance performance in time-series learning tasks \cite{SUZ22}. 

Despite these advantages, QRC faces a fundamental challenge when applied to sequential data. Each measurement collapses the quantum state, leading to a loss of quantum information \cite{FUJ17}. Reconstructing the system state therefore requires re-initializing the entire time series up to the current step, resulting in increasing computational overhead and circuit depth \cite{CIN24, YAS23, HU24, MUJ23, KOB24}. In noisy circuits, this results in noise accumulation, which further degrades performance.  
\begin{figure}[!t]
    \centering
    \includegraphics[width=\columnwidth, clip, trim=0 10pt 0 0]{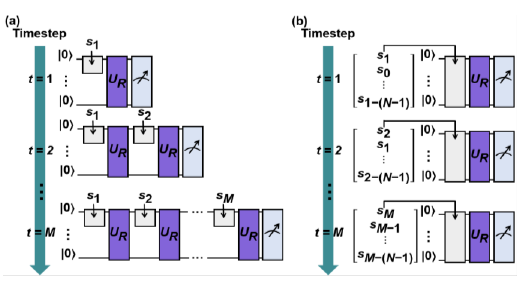}
    \caption{Conceptual Diagrams of QRC and TD-QELM.
    (a) In the restarting protocol, input encoding and unitary evolution $U_R$ are sequentially applied at each timestep to incorporate past information. This causes circuit depth and computational cost to grow quadratically with sequence length $M$.
    (b) The current input $s_t$ and $N-1$ time-delayed inputs ${s_{t}, s_{t-1}, \dots, s_{t-(N-1)}}$ are simultaneously encoded into $N$ qubits, keeping the circuit shallow regardless of $M$.}
    \label{fig:architecture}
\end{figure}
Several strategies have been proposed to mitigate these limitations. Mujal \textit{et al.} employed weak measurements, while \v{C}indrak \textit{et al.} exploited the fading-memory property of the reservoir to periodically reset the system \cite{MUJ23, CIN24}. Other studies have introduced feedback mechanisms or auxiliary qubits to counteract state collapse \cite{PFE22, KOB24, YAS23}. Although these techniques improve practicality, QRC still relies on sequential processing, handling one timestep at a time.  

To address these challenges, we propose the \emph{Time-Delayed Quantum Extreme Learning Machine} (TD-QELM), inspired by the classical \emph{Time-Delayed Extreme Learning Machine} (TD-ELM) \cite{BUT13b}. 

Motivated by Ref.~\cite{CIN24}, which shows that a quantum reservoir does not necessarily retain long input histories, TD-QELM encodes a fixed temporal window of delayed inputs, rather than extending the sequential QRC dynamics over increasingly long input histories. This design reformulates the quantum feature-map framework of QELM/QRC \cite{FUJ21, INN23, DEL25} for time-series prediction and yields quantum circuits whose depth is independent of the input sequence length. It thereby suppresses noise accumulation and reduces the computational complexity from quadratic to linear order.
The authors of \cite{XIO23} provided evidence that QELM architectures built from random circuits exhibit strong concentration effects, which severely limit their learning capability. To avoid these issues, our approach employs the transverse-field Ising model as a structured reservoir in the noiseless simulation. We then implement this reservoir via a one-step Trotterization of a 6-qubit model, resulting in shallow quantum circuits on IBM’s 127-qubit superconducting processor \textit{ibm\_kawasaki} as well as its corresponding simulator \texttt{FakeKawasaki} \cite{IBM25}.

The effectiveness of TD-QELM is evaluated using the Nonlinear AutoRegressive Moving Average (NARMA) benchmark for time-series forecasting. Our results show that TD-QELM consistently outperforms the standard QRC protocol across different noise conditions. In particular, TD-QELM maintains stable performance for longer input sequences, whereas QRC performance deteriorates due to increasing noise accumulation.
These results indicate that TD-QELM enables more robust and resource-efficient time-series forecasting on current NISQ devices.




\textit{Quantum reservoir computing.}
Quantum Reservoir Computing (QRC) aims to leverage the large Hilbert space of a quantum system  to perform time-series prediction tasks \cite{FUJ17}. 
The system is described by a Hamiltonian $H$ and its corresponding unitary operator $U_R = \exp(-iHT)$ for some $T > 0$.
The approach typically works by first encoding the current input $s_t$ of a time series $(s_t)_t$ into one of the qubits via
\begin{align}
    \ket{\psi_{\mathrm{in}}(s_t)} = \sqrt{1 - s_t}\, \ket{0} + \sqrt{s_t}\, \ket{1},
    \label{eq:enc}
\end{align}
which is applied to the first qubit. The full system state is then given by
\begin{align}
    \rho_{\mathrm{in}}(s_t) = \ket{\psi_{\mathrm{in}}(s_t)}\bra{\psi_{\mathrm{in}}(s_t)} \otimes \Tr_1\!\big(\rho(s_{t-1})\big),
\end{align}
where $\rho(s_{t-1})$ denotes the system state before encoding. After evolution under the reservoir $U_R$, the system state becomes
\begin{align}
    \rho(s_t) = U_R\, \rho_{\mathrm{in}}(s_t)\, U_R^\dagger.
\end{align}

In QRC, expectation values of the observables $\{\sigma_z^{(k)}\}_{k=1,\ldots,K}$ define the reservoir output features and are given by $x_{t,k} = \Tr\!\big(\sigma_z^{(k)} \rho(s_t)\big)$, where $K$ is the number of measured observables.
To increase the readout dimension, time-multiplexing at $N_V$ times $\tau_n = nT/N_V$ with $n\in \{1, 2, \ldots, N_V\}$ is employed. The multiplexed features $x_{t,(k,n)}$ yield $N_R = K N_V$ expectation values per input, forming $\mathbf{x}_t \in \mathbb{R}^{N_R}$ and the \textit{state matrix} $X \in \mathbb{R}^{T \times N_R}$.
\begin{align}
    y_t = \mathbf{x}_t \mathbf{w}_{\mathrm{out}}
\end{align}
, where the readout weights $\mathbf{w}_{\mathrm{out}}$ are optimized to minimize the loss $L = (\mathbf{y}-\hat{\mathbf{y}})^2$, and are obtained as
\begin{align}
    \mathbf{w}_{\mathrm{out}} = (X^\top X)^{-1} X^\top \hat{\mathbf{y}},
\end{align}
where $\mathbf{y}$ is the desired target vector. 
Shot noise scales as $1/\sqrt{N_{\mathrm{shots}}}$ and can act as a regularizer similarly to Tikhonov regularization \cite{BIS95}. We added Appendix.~\ref{app:error2}, showing this equivalence for QRC and TD-QELM.

\begin{figure}[t]
    \centering
    \includegraphics[scale=1.0]{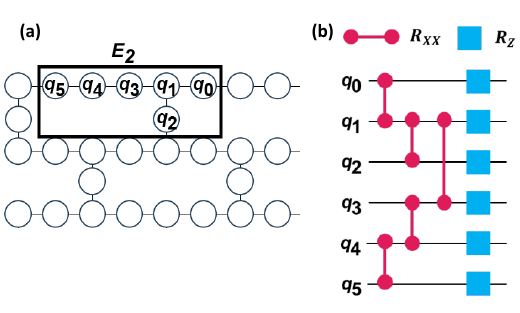}
    \caption{(a) Connectivity map of the qubits used in the experiment within the \textit{ibm\_kawasaki} device topology (partially shown), where each $q_i$ represents a qubit. The connections define a hardware-specific topology $E_2$. (b) Hardware-efficient circuit using Trotterization, designed to implement a multi-qubit quantum system under real-device connectivity constraints. This design reduces gate operations while maintaining sufficient entanglement on NISQ hardware.}
    \label{fig:topology}
\end{figure}
Since the quantum state collapses after each measurement, the system must be fully reinitialized at every input of the time series (see \cref{fig:architecture}(a)). The $s_n$-th input therefore requires $n_{\mathrm{enc}} = n$ encodings and $n_{R} = n$ reservoir evolutions. The computational cost of QRC for a time series of length $M$ is  
\begin{align}
    T_{\mathrm{QRC}} 
    &= \sum_{n=1}^M n
    = \frac{M(M+1)}{2} \in \mathcal{O}(M^2),
\label{eq:T_QRC}
\end{align}
demonstrating quadratic scaling \cite{FUJ17, CIN24}.

\textit{Time-Delayed QELM.}
To address the need for re-initializing the full time series, we propose a Quantum Extreme Learning Machine approach for time-series prediction, termed TD-QELM(see \cref{fig:architecture}(b)). In TD-QELM, the system is initialized in the state $\ket{\psi_{\mathrm{init}}} = \ket{00\ldots0}$ for each input $s_t$. As input to the TD-QELM, we consider a vector of $N$ previous inputs,
$
    \mathbf{s}_t = (s_{t-\tau_1}, s_{t-\tau_2}, \ldots, s_{t-\tau_N}).
$
After encoding, the resulting state is
$
    \ket{\psi_{\mathrm{in},t}} = U_E(\mathbf{s}_t)\, \ket{\psi_{\mathrm{init}}}.
$
In this work, each component of $\mathbf{s}_t$ is encoded into a separate qubit following Eq.~\eqref{eq:enc}. After evolution under $U_R$, the output state and density matrix are given by 
\begin{align}
    \ket{\psi(s_t)} = U_R\, \ket{\psi_{\mathrm{in},t}}, ~~~\rho(s_t) = \ket{\psi(s_t)}\bra{\psi(s_t)}.
\end{align}
Given that the number of qubits is $N$, this approach requires only $n_{\mathrm{enc}} = 1$ encoding of the input vector $\mathbf{s}_t$ and $n_{R} = 1$ reservoir evolution $U_R$. The computational cost for TD-QELM is therefore  
\begin{align}
    T_{\mathrm{TD\text{-}QELM}} 
    &= \sum_{n=1}^M 1
    = M 
    \in \mathcal{O}(M),
    \label{eq:T_QELM}
\end{align}
showing linear complexity with respect to the length of the time series.

\begin{table}[!b]
\centering
\caption{NMSE of TD-QELM and QRC for the NARMA10 task with varying readout dimension $N_R = N_V N_S$. The results of a linear regression model (LR) with readout dimension are shown for comparison. Boldface indicates the best NMSE for each readout dimension. Results are computed on a time series of length $M = 5000$, with 10\% used for washout, 70\% for training, and 20\% for testing.}
\label{tab:1}
\begin{tabular}{lccc}
\toprule
\multirow{2}{*}{\makecell[b]{Readout\\dim. $N_R$}} &$~~~~$ TD-QELM $~~~~$ & $~~~~$ QRC $~~~~$ & $~~~~$ LR $~~~~$ \\
  & $~~~~$ NMSE $~~~~$ & $~~~~$ NMSE $~~~~$ & $~~~~$ NMSE  $~~~~$  \\
\midrule
 $30$        & $9.83\cdot10^{-4}$ & $3.33\cdot10^{-3}$ & $\mathbf{7.47\cdot10^{-4}}$ \\
 $60$  & $\mathbf{4.55\cdot10^{-4}}$ & $2.73\cdot10^{-3}$ & $7.38\cdot10^{-4}$ \\
 $120$  & $\mathbf{3.08\cdot10^{-4}}$ & $2.06\cdot10^{-3}$ & $7.58\cdot10^{-4}$ \\
 $300$  & $\mathbf{2.68\cdot10^{-4}}$ & $1.85\cdot10^{-3}$ & $8.34\cdot10^{-4}$\\
 $600$  & $\mathbf{2.73\cdot10^{-4}}$ & $1.59\cdot10^{-3}$ & $1.04\cdot10^{-3}$ \\
\bottomrule
\end{tabular}

\end{table}

A standard benchmark for testing both the nonlinearity and memory capacity of a reservoir is the NARMA10 time-series prediction task, which is defined as
\begin{align}
y_{t+1} = 0.3 y_t + 0.05 y_t \sum_{d=0}^{9}y_{t-d} + 1.5 s_t s_{t-9} + 0.1,
\label{eq:narma}
\end{align}
where $s_t$ is uniformly distributed in $[0,0.2]$. The time-step $t$ input-encoding we employ for the TD-QELM is defined as
\begin{align}
\lvert \psi_{\mathrm{in},t} \rangle = &
\lvert \psi(s_t) \rangle \otimes 
\lvert \psi(s_{t-1}) \rangle \otimes 
\lvert \psi(s_{t-2}) \rangle \nonumber \\ &\otimes 
\lvert \psi(s_{t-9}) \rangle \otimes 
\lvert \psi(s_{t-10}) \rangle \otimes 
\lvert \psi(s_{t-11}) \rangle,
\label{eq:encoding}
\end{align}
which follows the information processing capacity analysis of Kubota \cite{KUB21, DAM12}. Such delayed input encoding has been shown to significantly increase task performance in classical reservoir computing \cite{LIN24, PIC25, OWE25}. The target at time-step $t$ is given by $y_{t+1}$ according to \cref{eq:narma}.
As a performance metric, we compute the signal Normalized Mean Squared Error (NMSE), as commonly used in QRC \cite{NAK19}, where $\mathbf{y} = (y_1, \ldots, y_T)$ and $\mathbf{y}^{\mathrm{target}} = (y_1^{\mathrm{target}}, \ldots, y_T^{\mathrm{target}})$ denote the output and target signals. The NMSE is defined as
\begin{align}
\mathrm{NMSE} = 
\frac{\sum_{t=1}^{T} (y_t - y_t^{\mathrm{target}})^2}
{\sum_{t=1}^{T} (y_t^{\mathrm{target}})^2}.
\end{align}
For comparison, in machine learning one typically employs a variance-normalized error measure
$
\mathrm{NMSE}_{\mathrm{var}}
=
{\sum_{t=1}^{T} (y_t - y_t^{\mathrm{target}})^2}/
{\sum_{t=1}^{T} \bigl(y_t - \mathbb{E}[\mathbf{y}]\bigr)^2}, 
$
where
$
\mathbb{E}[\mathbf{y}]
=
{T}^{-1}\sum_{t=1}^{T} y_t
$
denotes the expectation value of the target signal. The two errors are related as follows (see App.~\ref{app:error} for the derivation): 
$\mathrm{NMSE}_{\mathrm{var}} = \mathrm{NMSE} \left( 1 + \frac{\mathbb{E}[\mathbf{y}]^2}{\mathrm{var}[\mathbf{y}]} \right)$.

To test TD-QELM, we compare this method against standard QRC practices. 
The most commonly used quantum reservoir in QRC is the \textit{transverse-field Ising model}, given by
\begin{align}
H = \sum_{i,j \in E} J_{i,j}\, \sigma_x^{(i)} \sigma_x^{(j)} 
+ h\sum_{i} \, \sigma_z^{(i)},
\end{align}
where $J_{i,j}$ are sampled uniformly from the interval $[-5,\, 5]$ and the external field is set to $h=5$. For a statistical analysis, we consider ten realizations of such reservoirs with random \(J_{i,j}\) and plot the averaged performance. 
For the theoretical simulations, we allow interactions between all qubits, i.e., the connectivity matrix is $E_1 = \{(i,j) \mid i,j \in \{0,\ldots,N_S-1\}\}$, where $N_S = 6$ is the number of qubits.
To implement quantum reservoirs on actual hardware, and considering that the current generation of NISQ devices cannot support large circuit depths, we employ a simple one-step Trotterization, as illustrated schematically in \cref{fig:topology}(b) \cite{PRE18}. This results in the unitary circuit 
\begin{align}
U(\boldsymbol{\theta}) = 
\Biggl(\prod_{i=0}^{N_S} R_Z^{(i)}(2h)\Biggr)
\Biggl(\prod_{i,j=0}^{N_S} R_{XX}^{(i,j)}(2J_{i,j})\Biggr)
\label{eq:circuit}
\end{align}
proposed by Kandala \textit{et al.} \cite{KAN17,FRY23,IVA25}.
Simulations of both TD-QELM and QRC with connectivity $E_1$ were performed using the \texttt{aer\_simulator\_density\_matrix} backend in IBM Qiskit.


We then consider a quantum hardware platform with restricted qubit coupling topology, given by
$
E_2 = \{(0,1),~(1,2),~(1,3),~(3,4),~(4,5)\}.
$
These circuits are implemented on the 127-qubit NISQ device \textit{ibm\_kawasaki} (see \cref{fig:topology}(a)), and a corresponding backend simulation \texttt{FakeKawasaki} is also performed.

\begin{figure}[t]
    \centering
    \includegraphics[height=7.0cm]{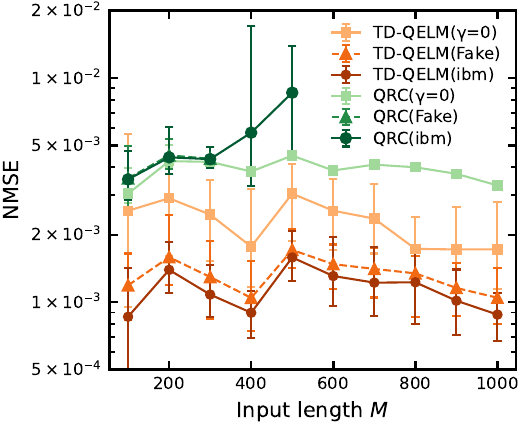}
  \caption{Average NMSE of TD-QELM and QRC as a function of input length $M$ under ideal noiseless simulation ($\gamma = 0$), a hardware-mimicking noise model (\texttt{FakeKawasaki}, denoted as ``Fake'' in the legend), and a real NISQ device (\textit{ibm\_kawasaki}, denoted as ``ibm''), using 10\% washout, 70\% training, and 20\% testing. $N_S = 6$ sites are considered with $N_V = 1$ multiplexing. TD-QELM maintains low NMSE even for long input sequences, whereas QRC shows rapid performance degradation. Error bars indicate the range between the maximum and minimum values over 10 trials.}
    \label{fig:result2}
\end{figure}

\textit{Results.} 
\Cref{tab:1} summarizes the NMSE for TD-QELM, QRC, and a linear model (LR). For TD-QELM and QRC, the NMSE values are averaged over 10 quantum systems. We use a 1-step Trotter circuit and measure the six sites in the Pauli-\(z\) direction, given by $\sigma_z^{(i)}$, a total of $N_V \in \{5, 10, 20, 50, 100\}$ times. For comparison, we consider the last $N_R$ inputs, $\mathbf{x}_t = (s_t, s_{t-1}, \ldots, s_{t-N_R+1})$, as the system state and construct a linear regression (LR) model. Training is performed in the same manner. We use $M = 5000$ data points in total, allocating 10\% for washout, 70\% for training, and 20\% for testing.

With a small readout dimension ($N_R = 30$), the best-performing model is the linear model (LR), which successfully captures the required memory aspects of the NARMA10 task. However, increasing the readout dimension allows the TD-QELM to sample more effectively from the large Hilbert space, thereby improving task performance and achieving an error of $\mathrm{NMSE}(\mathrm{TD\text{-}QELM}) = 2.73 \cdot 10^{-4}$, while the LR model stagnates due to the lack of nonlinearity in the input data, reaching only $\mathrm{NMSE}(\mathrm{LR}) = 7.38 \cdot 10^{-4}$.
QRC, on the other hand, exhibits errors roughly one order of magnitude larger than TD-QELM, with $\mathrm{NMSE}(\mathrm{QRC}) = 1.59 \cdot 10^{-3}$. This limitation arises from constant input overwriting, which prevents the system from accessing distant past inputs and performing the nonlinear operations required for the NARMA10 task.

Overall, these results suggest that, although TD-QELM has a restricted memory due to the encoding window, this limitation actually induces rich nonlinear features in the readout, thereby enhancing overall task performance, consistent with the findings reported in \cite{CIN24}. 
In Appendix~\ref{app:lorenz}, we show that TD-QELM significantly improves performance on chaotic time-series prediction tasks, using the Lorenz63 system as an example.





Next, we investigate both the QRC and TD-QELM protocols on actual quantum hardware, with a particular focus on noise accumulation. To this end, we examine how the input length affects the prediction performance of TD-QELM and QRC under three different conditions: (i) ideal noise-free simulation ($\gamma = 0$), (ii) simulation using the hardware-noise model \texttt{FakeKawasaki} from the Qiskit package, and (iii) execution on the NISQ device \textit{ibm\_kawasaki}, based on IBM's 127-qubit Eagle superconducting processor. The input length was varied from $M = 100$ to $M = 1000$ timesteps in increments of 100. In QRC, repeated re-initialization of the time series results in a computational cost that scales quadratically with the input length $M$ (see \cref{eq:T_QRC}). 
This quadratic overhead results in increased circuit depth on quantum hardware, making the system more susceptible to decoherence and gate errors \cite{CHE20b}. To test the influence of noise accumulation due to decoherence and gate errors, we adopt a simple QRC architecture without time multiplexing ($N_V = 1$).

As before, the initial 10\% of each sequence was used for washout, 70\% for training, and 20\% for testing. Each circuit was executed with $N_{\mathrm{shots}} = 8192$. Under these conditions, QRC could be executed up to 300 timesteps in simulation with \texttt{FakeKawasaki} and up to 500 timesteps on the real device (\textit{ibm\_kawasaki}), due to computational resource limitations.



\begin{figure}[t]
    \centering
    \includegraphics[height = 7.0cm]{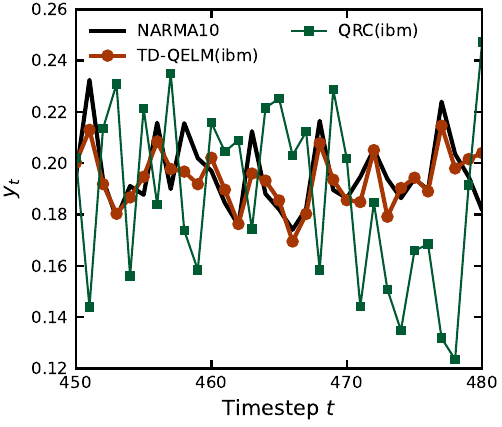}
    \caption{Time-series output for the NARMA10 task obtained on the \textit{ibm\_kawasaki} quantum hardware. The black curve shows the target signal (NARMA10), while the orange and green curves correspond to the outputs of TD-QELM (ibm) and QRC (ibm), respectively. }
    \label{fig:result1}
\end{figure}

\begin{table*}[!t]
\caption{Average NMSE of TD-QELM and QRC under different environments with varying input lengths. Each value represents the mean over 10 trials. The smallest NMSE value for each timestep is highlighted in bold.}
\label{tab:2}
\centering
\begin{tabular}{ccccccc}
\toprule
\multirow{2}{*}{Input length $M$} & \multicolumn{3}{c}{TD-QELM} & \multicolumn{3}{c}{QRC} \\
\cmidrule(lr){2-4} \cmidrule(lr){5-7}
 & Aer(Noiseless) & \texttt{FakeKawasaki} & \textit{ibm\_kawasaki} & Aer(Noiseless) & \texttt{FakeKawasaki} & \textit{ibm\_kawasaki} \\
\midrule
100  & $2.56\cdot10^{-3}$ & $1.19\cdot10^{-3}$ & $\mathbf{8.61\cdot10^{-4}}$ & $3.05\cdot10^{-3}$ & $3.57\cdot10^{-3}$ & $3.53\cdot10^{-3}$ \\
300  & $2.46\cdot10^{-3}$ & $1.30\cdot10^{-3}$ & $\mathbf{1.08\cdot10^{-3}}$ & $4.23\cdot10^{-3}$ & $4.37\cdot10^{-3}$ & $4.35\cdot10^{-3}$ \\
500  & $3.05\cdot10^{-3}$ & $1.71\cdot10^{-3}$ & $\mathbf{1.58\cdot10^{-3}}$ & $4.51\cdot10^{-3}$ & -- & $8.61\cdot10^{-3}$ \\
1000 & $1.72\cdot10^{-3}$ & $1.05\cdot10^{-3}$ & $\mathbf{8.81\cdot10^{-4}}$ & $3.32\cdot10^{-3}$ & -- & -- \\
\bottomrule
\end{tabular}
\end{table*}  
\Cref{fig:result2} illustrates the relationship between input length and prediction performance for TD-QELM (orange) and QRC (green) under different environments. 
The hardware execution times corresponding to the real quantum device in \mbox{\cref{fig:result2}} are reported in Appendix~\mbox{\ref{app:time}}, showing the quadratic and linear scaling according to Eq.~\ref{eq:T_QRC} for QRC and Eq.~\ref{eq:T_QELM} for TD-QELM.
In the noiseless simulation, both TD-QELM (light orange) and QRC (light green) remain largely constant with respect to input length. However, when executed on real IBM quantum devices, QRC's NMSE (dark green) increases substantially for longer inputs. This behavior is mainly attributed to the repeated re-initialization of the full time series, which increases the total number of circuit executions. In the presence of noise on current quantum hardware, this results in performance degradation, as evidenced by the significantly increased NMSE for $M > 400$. TD-QELM (dark orange), in contrast, maintains stable performance with respect to the input length $M$, since each input requires the same number of quantum circuits. This behavior is also accurately reproduced by the \texttt{FakeKawasaki} noise model. 
\Cref{tab:2} summarizes the average NMSE values corresponding to \cref{fig:result2} (see Appendix~\ref{app:std} for the standard deviations). The lowest NMSE for all input lengths and environments is obtained when TD-QELM is implemented on the real device. 
Here, \texttt{FakeKawasaki} and \textit{ibm\_kawasaki} exhibit comparable behavior, with \textit{ibm\_kawasaki} outperforming both the noisy simulation and the noiseless TD-QELM case. This noise-enhanced performance has been observed in several QRC studies, where intrinsic hardware noise can positively impact prediction accuracy \cite{SUZ22, KUB23, SAN24, FRA24}. In the case of TD-QELM, this implies that natural quantum noise present on NISQ devices can play a constructive role in improving predictive accuracy, underscoring the potential of noise-assisted quantum machine learning. 
To illustrate the qualitative prediction behavior, 
\mbox{\cref{fig:result1}}
shows a segment of the predicted NARMA10 waveform for a given input sequence. Comparison with the NARMA10 target sequence (black) reveals that QRC (dark green) fails to capture the task dynamics, whereas TD-QELM (dark orange), even with a limited readout dimension, closely follows the fluctuations and temporal structure of the target.
These results suggest that the repeated re-initialization of the full time series in QRC increases the total number of circuit executions, leading to effective noise accumulation and making QRC challenging to implement on current NISQ hardware. In contrast, TD-QELM not only exhibits increased robustness to noise but also achieves improved task performance, which can be attributed to its ability to map fewer data points onto highly nonlinear features.


\textit{Conclusion.}
This work examined limitations of quantum reservoir computing (QRC), particularly noise accumulation and the quadratic scaling of computational cost with input sequence length. To address these issues, we introduced the \emph{time-delayed quantum extreme learning machine} (TD-QELM), which enables parallel encoding of past inputs and reduces the time complexity to linear while improving task performance.\\
We first considered a transverse-field Ising model with all-to-all connectivity and implemented the reservoir evolution using a single-step Trotterization, resulting in shallow circuits suitable for noisy intermediate-scale quantum (NISQ) devices. Simulations, showed that TD-QELM consistently outperforms conventional QRC in task performance on the NARMA10 task, which is due to the rich non-linear features exhibited by restricted memory, also observed in \cite{CIN24}.\\
To validate these findings under realistic conditions, we conducted simulations with \texttt{FakeKawasaki} and performed experiments on the \textit{ibm\_kawasaki} quantum processor. The hardware topology required Hamiltonians tailored to the native qubit connectivity, enabling the design of hardware-constrained quantum reservoirs.
Across noiseless simulations, noisy simulations, and hardware execution, TD-QELM outperforms QRC in the NARMA10 task.
The need to re-initialize the full time series on current noisy quantum circuits leads to noise accumulation, making QRC difficult to scale. TD-QELM, on the other hand, maintains stable performance for longer sequences due to its shallow-circuit design, reduced operation count, and linear scaling. When considering realistic hardware, TD-QELM performs better in the presence of noise than under noiseless conditions, consistent with previously reported noise-assisted reservoir behavior \cite{SUZ22, KUB23, SAN24, FRA24}.  

Overall, this work demonstrates TD-QELM as a scalable and hardware-efficient alternative to conventional QRC. By mitigating noise accumulation, improving prediction accuracy, and remaining compatible with NISQ devices, TD-QELM provides a promising pathway toward practical quantum time-series processing. 

\textit{Acknowledgements.} The authors would like to express their sincere gratitude to Dr. Atsushi Matsuo and Mr. Toru Imai of IBM Japan for their invaluable discussions and constructive suggestions.

\bibliography{lit_aig}

\appendix

\section{Relationship between error definitions}\label{app:error}
{In this appendix, we derive the relationship between the signal-energy normalized NMSE commonly used in quantum reservoir computing and the variance-normalized 
$\mathrm{NMSE}_{\mathrm{var}}$
frequently employed in machine learning. The derivation makes explicit how the two error measures differ due to the contribution of the mean (DC component) of the target signal.}

\begin{align}
    \mathrm{NMSE} = 
\frac{\sum_{t=1}^{T} (y_t - y_t^{\mathrm{target}})^2}
{\sum_{t=1}^{T} (y_t^{\mathrm{target}})^2},
\\
\mathrm{NMSE}_{\mathrm{var}} =
\frac{\sum_{t=1}^{T} (\hat y_t - y_t)^2}
{\sum_{t=1}^{T} \bigl(y_t - \mathbb{E}[\mathbf{y}]\bigr)^2}.
\end{align}

We start from the variance-normalized definition and multiply by unity:
\begin{align}
\mathrm{NMSE}_{\mathrm{var}}
&=
\mathrm{NMSE}_{\mathrm{var}}
\frac{\sum_{t=1}^{T} y_t^2}{\sum_{t=1}^{T} y_t^2} \nonumber \\
&=
\mathrm{NMSE}
\frac{\sum_{t=1}^{T} y_t^2}
{\sum_{t=1}^{T} \bigl(y_t - \mathbb{E}[\mathbf{y}]\bigr)^2}.
\end{align}

To relate the denominators, we use the identity
\begin{align}
\sum_{t=1}^{T} \bigl(y_t - \mathbb{E}[\mathbf{y}]\bigr)^2
&=
\sum_{t=1}^{T} y_t^2
- 2\,\mathbb{E}[\mathbf{y}] \sum_{t=1}^{T} y_t
+ T\,\mathbb{E}[\mathbf{y}]^2 \nonumber \\
&=
\sum_{t=1}^{T} y_t^2
- 2T\,\mathbb{E}[\mathbf{y}]^2
+ T\,\mathbb{E}[\mathbf{y}]^2 \nonumber \\
&=
\sum_{t=1}^{T} y_t^2
- T\,\mathbb{E}[\mathbf{y}]^2,
\end{align}
which can be rearranged as
\begin{align}
\sum_{t=1}^{T} y_t^2
=
\sum_{t=1}^{T} \bigl(y_t - \mathbb{E}[\mathbf{y}]\bigr)^2
+ T\,\mathbb{E}[\mathbf{y}]^2.
\end{align}

Substituting this expression back yields
\begin{align}
\mathrm{NMSE}_{\mathrm{var}}
&=
\mathrm{NMSE}
\frac{\sum_{t=1}^{T} \bigl(y_t - \mathbb{E}[\mathbf{y}]\bigr)^2
+ T\,\mathbb{E}[\mathbf{y}]^2}
{\sum_{t=1}^{T} \bigl(y_t - \mathbb{E}[\mathbf{y}]\bigr)^2}
\nonumber \\
&=
\mathrm{NMSE}
\left(
1
+
\frac{\mathbb{E}[\mathbf{y}]^2}
{\frac{1}{T}\sum_{t=1}^{T} \bigl(y_t - \mathbb{E}[\mathbf{y}]\bigr)^2}
\right)
\nonumber \\
&=
\mathrm{NMSE}
\left(
1 + \frac{\mathbb{E}[\mathbf{y}]^2}{\mathrm{var}[\mathbf{y}]}
\right).
\end{align}

\section{Lorenz task}\label{app:lorenz}
In this appendix, we present additional simulation results for the Lorenz task. The Lorenz dynamics are given by
\begin{equation}
\displaystyle
\begin{aligned}
    \dot{x} &= \sigma(y-x),\\
    \dot{y} &= x(\rho-z)-y,\\
    \dot{z} &= xy-\beta z,
\end{aligned}
\label{eq:lorenz}
\end{equation}
where we use the parameters $\sigma=10$, $\rho=28$, and $\beta=8/3$.
We integrate the system with $dt = 0.001$ and discretize it using $\Delta t = 0.1$ to construct the time series $x(n\Delta t)=x_n$, $y(n\Delta t)=y_n$, and $z(n\Delta t)=z_n$.
Using $x_n$ as the input signal, we evaluate two prediction tasks: the self-prediction task $x_n \to x_{n+1}$ and the cross-prediction task $x_n \to z_n$.
The input and target time series are rescaled to the interval $[0,1]$.
The results for the Lorenz $x_n \to x_{n+1}$ and $x_n \to z_n$ tasks are summarized in Table~\mbox{\ref{tab:lorenz_results}}.
For both tasks, TD-QELM yields lower NMSE values than QRC for most readout dimensions.
These results demonstrate that TD-QELM can achieve significant performance improvements: up to one order of magnitude for the Lorenz $x_n \to x_{n+1}$ task, and only limited improvements for small readout dimensions in the Lorenz $x_n \to z_n$ task.
This behavior can be attributed to the nonlinear feature response of TD-QELM. The Lorenz $x_n \to x_{n+1}$ task requires less memory and stronger nonlinearity, whereas the Lorenz $x_n \to z_n$ task exhibits slower dynamics, increasing the memory requirements. In this case, the significantly enhanced nonlinearity of the system for small readout dimensions leads to improved task performance. However, increasing the readout dimension allows the reservoir to sample a richer input history, thereby enabling better computation and reducing the relative advantage of TD-QELM.

\begin{table}[t]
\centering
\caption{
Lorenz prediction results.
All NMSE values are reported in units of $10^{-3}$.
The best performance is highlighted in bold.
The last column reports the improvement factor 
$\mathrm{NMSE}_{\mathrm{QRC}}/\mathrm{NMSE}_{\mathrm{TD\text{-}QELM}}$.
}
\label{tab:lorenz_results}

\renewcommand{\arraystretch}{1.2}
\setlength{\tabcolsep}{7pt}

\begin{tabular}{@{}lcccc@{}}
\toprule
\multicolumn{5}{@{}l}{(a) Lorenz $x_n \to x_{n+1}$ task} \\
\midrule
\makecell[b]{Readout\\dim. $N_R$}
& \makecell[b]{TD-QELM\\NMSE}
& \makecell[b]{QRC\\NMSE}
& \makecell[b]{LR\\NMSE}
& Improvement \\
\midrule
$30$  & $\mathbf{2.73}$  & $8.12$ & $67.5$ & $2.97$ \\
$60$  & $\mathbf{0.972}$ & $3.06$ & $67.1$ & $3.15$ \\
$120$ & $\mathbf{0.184}$ & $1.61$ & $69.7$ & $8.75$ \\
$300$ & $\mathbf{0.154}$ & $1.19$ & $79.8$ & $7.73$ \\
\midrule
\multicolumn{5}{@{}l}{(b) Lorenz $x_n \to z_n$ task} \\
\midrule
\makecell[b]{Readout\\dim. $N_R$}
& \makecell[b]{TD-QELM\\NMSE}
& \makecell[b]{QRC\\NMSE}
& \makecell[b]{LR\\NMSE}
& Improvement \\
\midrule
$30$  & $\mathbf{0.696}$ & $2.27$  & $141$ & $3.26$ \\
$60$  & $\mathbf{0.154}$ & $0.401$ & $142$ & $2.60$ \\
$120$ & $\mathbf{0.0503}$ & $0.101$ & $140$ & $2.01$ \\
$300$ & $0.0450$ & $\mathbf{0.0434}$ & $155$ & $0.96$ \\
\bottomrule
\end{tabular}
\end{table}

\section{Hardware execution time on IBM quantum devices}\label{app:time}
\begin{figure}[t]
    \centering
    \includegraphics[height=7.0cm]{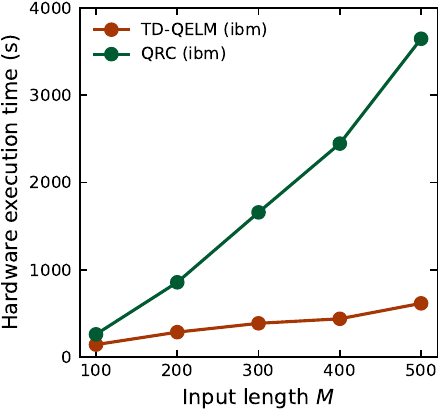}
  \caption{
  Hardware execution times of TD-QELM and QRC corresponding to the real-device experiments in \mbox{\cref{fig:result2}}.}
    \label{fig:result5}
\end{figure}

In this appendix, we report the hardware execution times corresponding to the real-device results obtained on \textit{ibm\_kawasaki} and shown in \mbox{\cref{fig:result2}}.
The reported time denotes the quantum-circuit execution time on the quantum device and does not include queueing or job-waiting time on the IBM Quantum platform.
\mbox{\Cref{fig:result5}} 
compares the hardware execution times of QRC and TD-QELM for input lengths of $M=100, 200, 300, 400,$ and $500$ under the same experimental setting.
As the input length increases, the execution time of TD-QELM grows approximately linearly, whereas that of QRC increases much more rapidly, consistent with the quadratic scaling expected from the repeated re-initialization protocol.
For example, when the input length is increased from $M=100$ to $M=500$, the execution time increases by approximately a factor of 4.3 for TD-QELM, whereas it increases by approximately a factor of 14 for QRC.
These results suppor the expected difference in hardware-execution scaling and indicate that TD-QELM is more hardware-efficient than QRC for the input lengths considered.

\section{Standard deviation of NMSE over repeated trials}\label{app:std}

\begin{table*}[!t]
\caption{Standard devitation of NMSE of TD-QELM and QRC under different environments with varying input length. Each value represents the standard deviation over 10 independent trials. Boldface indicates the smallest standard deviation for each input length.}
\label{tab:3}
\centering
\begin{tabular}{ccccccc}
\toprule
\multirow{2}{*}{Input length $M$} & \multicolumn{3}{c}{TD-QELM} & \multicolumn{3}{c}{QRC} \\
\cmidrule(lr){2-4} \cmidrule(lr){5-7}
     & Aer(Noiseless) & \texttt{FakeKawasaki} & \textit{ibm\_kawasaki}
     & Aer(Noiseless) & \texttt{FakeKawasaki} & \textit{ibm\_kawasaki}\\
     \midrule
     100 & $1.48\cdot10^{-3}$ & $2.86\cdot10^{-4}$ & $\mathbf{2.85\cdot10^{-4}}$
         & $3.68\cdot10^{-4}$ & $7.74\cdot10^{-4}$ & $6.30\cdot10^{-4}$ \\
     300 & $7.30\cdot10^{-4}$ & $3.33\cdot10^{-4}$ & $2.18\cdot10^{-4}$
         & $\mathbf{1.25\cdot10^{-4}}$ & $2.78\cdot10^{-4}$ & $3.32\cdot10^{-4}$ \\
     500 & $8.46\cdot10^{-4}$ & $2.41\cdot10^{-4}$ & $2.79\cdot10^{-4}$
         & $\mathbf{7.67\cdot10^{-5}}$ & -- & $3.36\cdot10^{-3}$ \\
     1000 & $5.55\cdot10^{-4}$ & $2.11\cdot10^{-4}$ & $1.39\cdot10^{-4}$
         & $\mathbf{1.21\cdot10^{-5}}$ & -- & -- \\
\bottomrule
\end{tabular}

\end{table*}

In this appendix, we quantify the variability of the NMSE values reported in \Cref{tab:2} of the main text by evaluating the standard deviation over independent trials. The standard deviation of the NMSE is calculated as
\begin{equation}
    \sigma_{\mathrm{NMSE}}
    = 
    \sqrt{
    \frac{1}{N-1}
    \sum_{i=1}^{N}
    \left(
    \mathrm{NMSE}_i
    -
    \overline{\mathrm{NMSE}}
    \right)^2
    },
    \label{eq:std_nmse}
    \end{equation}
where $\mathrm{NMSE}_i$ denotes the NMSE obtained in the $i$th independent trials, $\overline{\mathrm{NMSE}}$ is the mean NMSE over $N$ trial, and $N=10$.
\Cref{tab:3} summarizes the standard deviation of the NMSE values evaluated over independent trials. In particular, for the experimental results obtained on quantum hardware, TD-QELM exhibits smaller standard deviations than QRC, indicating more stable and invariant prediction performance with respect to variations in the initial conditions.

\section{Relationship between Tikhonov regularization and regularization by noise}
\label{app:error2}

In this appendix, we briefly discuss the relationship between Tikhonov regularization and regularization by noise for a linear readout layer.
In particular, we show that the connection considered here can be understood as a simple special case of the more general analysis presented in Ref.~\mbox{\cite{BIS95}}.

\textbf{Tikhonov regularization}
\\
Tikhonov regularization augments the squared-loss objective with a quadratic penalty on the model parameters:

{
\begin{align}
\displaystyle
\mathcal{L}_{\mathrm{Tik}}(w)
=
\mathbb{E}[(y-w^\top x)^2]
+
\lambda \|Aw\|_2^2 ,
\end{align}
}
where $\lambda > 0$ controls the regularization strength and $A$ defines the structure of the penalty.
Using
\begin{equation}
\displaystyle
\|Aw\|_2^2 = (Aw)^\top(Aw) = w^\top A^\top A w,
\end{equation}
the objective can be written as
\begin{equation}
\displaystyle
\mathcal{L}_{\mathrm{Tik}}(w)
=
\mathbb{E}[(y-w^\top x)^2]
+
w^\top (\lambda A^\top A) w.
\label{eq:Tikhanov}
\end{equation}
In the special case $A = I$, this reduces to ridge regularization:
\begin{equation}
\displaystyle
\mathcal{L}_{\mathrm{ridge}}(w)
=
\mathbb{E}[(y-w^\top x)^2]
+
\lambda \|w\|_2^2 .
\end{equation}

\textbf{Regularization by noise}
\\
We now show that training with noisy features induces a regularization effect of the same form. Let the measured feature vector be

\begin{align}
\tilde{x} = x + \epsilon,
\end{align}

where $\epsilon$ denotes zero-mean noise arising from finite-shot measurements, with covariance $\mathbb{E}[\epsilon \epsilon^\top] = \Sigma_\epsilon$.
We consider a linear model $y \approx w^\top \tilde{x}$ trained with squared loss.
The expected loss is
\begin{equation}
\displaystyle
\mathbb{E}_{x,y,\epsilon}\big[(y - w^\top \tilde{x})^2\big]
=
\mathbb{E}_{x,y,\epsilon}\big[(y - w^\top (x+\epsilon))^2\big].
\end{equation}
Expanding the square,

\begin{equation}
\displaystyle
\begin{aligned}
(y - w^\top x - w^\top \epsilon)^2
&=
(y - w^\top x)^2 \\
&\quad - 2(y - w^\top x)(w^\top \epsilon)
+ (w^\top \epsilon)^2.
\end{aligned}
\end{equation}

Taking expectation over the noise $\epsilon$, and assuming unbiased shot noise, i.e., $\mathbb{E}_{\epsilon}[\epsilon \mid x,y]=0$, the cross term vanishes since
\begin{equation}
\displaystyle
\mathbb{E}_{\epsilon}\!\left[(y - w^\top x)(w^\top \epsilon)\mid x,y\right] = 0.
\end{equation}
Thus,
\begin{equation} 
\begin{aligned}
\mathbb{E}_{x,y,\epsilon}[(y - w^\top (x+\epsilon))^2]
&=
\mathbb{E}_{x,y}[(y - w^\top x)^2] \\
&\quad+
\mathbb{E}_{\epsilon}[(w^\top \epsilon)^2].
\end{aligned}
\end{equation}

The remaining term evaluates to
\begin{equation}    
\displaystyle
\mathbb{E}_{\epsilon}[(w^\top \epsilon)^2]
=
w^\top \mathbb{E}[\epsilon \epsilon^\top] w
=
w^\top \Sigma_\epsilon w.
\end{equation}
Therefore,
\begin{equation}
\displaystyle
\mathbb{E}[(y - w^\top (x+\epsilon))^2]
=
\mathbb{E}[(y - w^\top x)^2]
+
w^\top \Sigma_\epsilon w.
\label{eq:noise}
\end{equation}
With the identification $\Sigma_\epsilon = \lambda A^\top A$, regularization by noise \mbox{(\cref{eq:noise})} and Tikhonov regularization \mbox{(\cref{eq:Tikhanov})} take the same form.
In particular, for isotropic noise, $\Sigma_\epsilon = \sigma^2 I$, this reduces to ridge regularization.


\end{document}